# BOUND STATE SOLUTIONS OF THE GENERALIZED SHIFTED HULTHEN POTENTIAL

C. O. Edet[1]; P. O. Okoi[2]; A. S. Yusuf[3] and P.O. Ushie[4]

[1]Theoretical Physics Group, Department of Physics, University of Port Harcourt, Rivers State, Nigeria.

[2]Department of Physics, University of Calabar, Etagbor, PMB 1115, Calabar, Cross River State, Nigeria.

[3]Department of Physics, Federal University of Technology, PMB 65, Gidan Kwanu, Minna-Bida Road, Niger State, Nigeria.

[4]Department of Physics, Cross River University of Technology, Calabar, Nigeria.

## ABSTRACT

In this study, we obtain an approximate solution of the Schrödinger equation in arbitrary dimensions for the generalized shifted Hulthén potential model within the framework of the Nikiforov-Uvarov method. The bound state energy eigenvalues were computed and the corresponding eigenfunction was also obtained. It is found that the numerical eigenvalues were in good agreement for all three approximations scheme used. Special cases were considered when the potential parameters were altered, resulting into Hulthén Potential and Woods-Saxon Potential respectively. Their energy eigenvalues expressions agreed with the already existing literatures. A straightforward extension to the s-wave case for Hulthén potential and Woods-Saxon Potential cases are also presented.

## 1. INTRODUCTION

In the last two decades, theoretical physicists have made unprecedented progress in the study of the behavior of different quantum mechanical systems[1]. Apparently, this progress has been made possible by obtaining exact or approximate solutions of the nonrelativistic and relativistic wave equations for different physical potentials of interest. The exact or approximate solutions of these equations with central potentials play a crucial role in quantum mechanics [2-5].

*Corresponding author, E-mail: collinsokonedet@gmail.com

The analytical solution of Schrödinger equation with $\ell = 0$ and $\ell \neq 0$ for some physical potentials has been addressed by many researchers. Some of these exponential-type potentials include, Manning-Rosen potential [6-9], Eckart potential [10-12], Poschl Teller Like Potential [13-14], a hyperbolic potential [15-17], generalized Morse potential [18], the Morse potential [19] and screen Coulomb potential[20].

The radial Schrödinger equation for these potentials can be solved exactly for $\ell = 0$ (s-wave) but cannot be solved for these potentials for $\ell \neq 0$. To obtain the solution for $\ell \neq 0$, we employ the Pekeris-type approximation scheme to deal with the centrifugal term or solve numerically [21]. The most widely used approximation was introduced by Pekeris [22] and another form was suggested by Greene and Aldrich [23] and Qiang et al [24].

Several methods have been employed to obtain the solutions of the nonrelativistic wave equations with a chosen potential model. These includes the Nikiforov-Uvarov method (NU) [25-28], Qiang-Dong proper quantization rule [29], Factorization Method [30-32], Supersymmetry Quantum Mechanics (SUSYQM) [33-35], Asymptotic Iteration Method (AIM) [36,37], algebraic approach [38] etc.

The Hulthén potential [39, 40] plays a vital role in atomic and molecular physics. [41] It has also been used to explain the electronic properties of some alkali halides. [42] More so, it resembles the Coulomb interaction in structure. The Hulthén potential is one of the important short-range potentials(i.e., large $b$ and small $l$[43]) in physics. The potential has been used in nuclear and particle physics, atomic physics, solid state physics and its bound state and scattering properties have been investigated by employing numerous techniques. General wave functions of this potential have been used in solid-state and atomic physics problems. It should be noted that Hulthén potential is a special case of Eckart potential.

The shifted Hulthén potential has not received great attention from researchers. The Dirac equation with this potential has been investigated [44]. Recently, Ikot *et al*. (2015) used SUSYQM approach to solve the Dirac equation with this potential in the presence of the Yukawa-like tensor (YLT) and generalized tensor (GLT) interactions [45]. Ikot *et al*. (2013) obtained the approximate analytical solutions of the Dirac equation for this potential within the framework of spin and pseudospin symmetry limits for arbitrary spin–orbit quantum number $k$ using the supersymmetry quantum mechanics [46].

The Shifted Hulthén potential is given as [44-46];

$$V(r) = \frac{(V_0 + 1/2b^2)e^{-r/b}}{1 - e^{-r/b}} + \frac{V_1 e^{-2r/b}}{(1 - e^{-r/b})^2} \tag{1}$$

which differs from the special potential Hulthén potentials [44] by the second term on the right-hand side. In the potential relation, $b$ is the range of the potential, $V_0$ and $V_1$ represents the depth of the potential well[44]. If $V_1 = 0, (V_0 + 1/2b^2) = -(V_0 + 1/2b^2)$, Eq. (1) reduces to the special Hulthén Potential [46], this is also different from the usual Hulthén potential with a new term $-1/2b^2$.

Motivated by the success in obtaining analytical solution of the Dirac Equation (Relativistic Quantum Mechanics) with this potential Eq. (1) using the standard method by Jian *et al.* [44] and Supersymmetric Quantum Mechanics(SUSYQM) method by Ikot *et al* [45-46]. We attempt to modify the shifted Hulthén Potential by introducing a deformation parameter($q$), and solve for this potential using a different method called the Nikiforov-Uvarov method (N-U) [47]. The essence of introducing the deformation parameter is to have a wider range of applications. The proposed potential (generalized shifted Hulthén potential) is given by;

$$V_q(r) = \frac{(V_0 + 1/2b^2)e^{-r/b}}{1 - qe^{-r/b}} + \frac{V_1 e^{-2r/b}}{(1 - qe^{-r/b})^2} \qquad (2)$$

In the potential relation, $b$ is the range of the potential and $V_0$ represents the depth of the potential well. Bearing in mind the outcomes of the Kratzer potential in the adhoc inverse square term for small distances [46].

The short range generalized shifted Hulthén potential will be solved within the framework of the Pekeris type approximations suggested by [24] to solve the Schrödinger equation(Non-relativistic Quantum Mechanics) for any arbitrary $\ell$ −state. These approximations are [32, 48, 49]:

$$\frac{1}{r^2} \approx \frac{e^{-r/b}}{b^2(1 - qe^{-r/b})^2} \qquad (3)$$

eq (3) is the commonly used approximation [24];

$$\frac{1}{r^2} \approx \frac{1}{b^2}\left[\frac{e^{(1-r)/b}}{(1 - qe^{-r/b})} + \frac{e^{-2r/b}}{(1 - qe^{-r/b})^2}\right] \qquad (4)$$

and the one suggested by [24];

$$\frac{1}{r^2} \approx \frac{1}{b^2}\left[\frac{1}{12} + \frac{e^{-r/b}}{(1 - qe^{-r/b})^2}\right] \qquad (5)$$

Eqs. (4) and (5) are more general than Eqs. (3), (4) and (5) give a better approximation to the centrifugal term when b is small [24].

In view of the above, the research reported in the present paper was also motivated by the fact that the non-relativistic treatment of the shifted Hulthén potential have not been reported in the available literature.

This paper is organized as follows. In section 2, the review of the Nikiforov-Uvarov Method is presented. In section 3, this method is applied to solve the radial Schrödinger equation with the generalized shifted Hulthén potential. In section 4, numerical calculations are given, the results are compared for the three approximations understudy and we discuss the results. In section 5, special cases are presented and comments are made and we give a brief concluding remark in section 6.

## 2. REVIEW OF NIKIFOROV-UVAROV METHOD

The Nikiforov-Uvarov (NU) method is based on solving the hypergeometric-type second-order differential equations by means of the special orthogonal functions. The main equation which is closely associated with the method is given in the following form [43]

$$\psi''(s) + \frac{\tilde{\tau}(s)}{\sigma(s)}\psi'(s) + \frac{\tilde{\sigma}(s)}{\sigma^2(s)}\psi(s) = 0 \tag{6}$$

Where $\sigma(s)$ and $\tilde{\sigma}(s)$ are polynomials at most second-degree, $\tilde{\tau}(s)$ is a first-degree polynomial and $\psi(s)$ is a function of the hypergeometric-type.

The exact solution of Eq. (6) can be obtained by using the transformation

$$\psi(s) = \phi(s)y(s) \tag{7}$$

This transformation reduces Eq. (6) into a hypergeometric-type equation of the form

$$\sigma(s)y''(s) + \tau(s)y'(s) + \lambda y(s) = 0 \tag{8}$$

The function $\phi(s)$ can be defined as the logarithm derivative

$$\frac{\phi'(s)}{\phi(s)} = \frac{\pi(s)}{\sigma(s)} \tag{9}$$

where $\pi(s) = \frac{1}{2}[\tau(s) - \tilde{\tau}(s)]$ \hfill (10)

with $\pi(s)$ being at most a first-degree polynomial. The second $\psi(s)$ being $y_n(n)$ in Eq. (7), is the hypergeometric function with its polynomial solution given by Rodrigues relation

$$y^{(n)}(s) = \frac{B_n}{\rho(s)}\frac{d^n}{ds^n}[\sigma^n \rho(s)] \tag{11}$$

Here, $B_n$ is the normalization constant and $\rho(s)$ is the weight function which must satisfy the condition

$$(\sigma(s)\rho(s))' = \sigma(s)\tau(s) \tag{12}$$

$$\tau(s) = \tilde{\tau}(s) + 2\pi(s) \tag{13}$$

It should be noted that the derivative of $\tau(s)$ with respect to $s$ should be negative. The eigenfunctions and eigenvalues can be obtained using the definition of the following function $\pi(s)$ and parameter $\lambda$, respectively:

$$\pi(s) = \frac{\sigma'(s) - \tilde{\tau}(s)}{2} \pm \sqrt{\left(\frac{\sigma'(s) - \tilde{\tau}(s)}{2}\right)^2 - \tilde{\sigma}(s) + k\sigma(s)} \tag{14}$$

where $k = \lambda - \pi'(s)$ \hfill (15)

The value of $k$ can be obtained by setting the discriminant of the square root in Eq. (9) equal to zero. As such, the new eigenvalue equation can be given as

$$\lambda_n = -n\tau'(s) - \frac{n(n-1)}{2}\sigma''(s), n = 0,1,2,\ldots \tag{16}$$

## 3 Bound State Solution

The radial Schrodinger equation in $D$ dimension can be given as [50]:

$$\left[\frac{d^2 R_{nl}}{dr^2} - \frac{2\mu V_q(r)}{\hbar^2} - \frac{(D+2\ell-1)(D+2\ell-3)}{4r^2} + \frac{2\mu E_{nl}}{\hbar^2}\right] R_{nl}(r) = 0 \qquad (17)$$

where $\mu$ is the reduced mass, $E_{nl}$ is the energy spectrum, $\hbar$ is the reduced Planck's constant and $n$ and $l$ are the radial and orbital angular momentum quantum numbers respectively (or vibration-rotation quantum number in quantum chemistry). Substituting Eq. (2) into Eq. (17) gives:

$$\left[\frac{d^2 R_{nl}}{dr^2} - \frac{2\mu}{\hbar^2}\left(\frac{(V_0+1/2b^2)e^{\frac{r}{b}}}{1-qe^{\frac{r}{b}}} + \frac{V_1 e^{\frac{2r}{b}}}{\left(1-qe^{\frac{r}{b}}\right)^2}\right) - \frac{(D+2\ell-1)(D+2\ell-3)}{4r^2} + \frac{2\mu E_{nl}}{\hbar^2}\right] R_{nl}(r) = 0 \qquad (18)$$

Simplifying further equation 18 becomes;

$$\left[\frac{d^2 R_{nl}}{dr^2} - \frac{2\mu}{\hbar^2}\left(\frac{(V_0+1/2b^2)e^{\frac{r}{b}}}{1-qe^{\frac{r}{b}}} + \frac{V_1 e^{\frac{2r}{b}}}{\left(1-qe^{\frac{r}{b}}\right)^2}\right) - \frac{(D+2\ell-1)(D+2\ell-3)}{4}\left(\frac{\alpha^2 e^{-\alpha r}}{(1-qe^{-\alpha r})^2}\right) + \frac{2\mu E_{nl}}{\hbar^2}\right] R_{nl}(r) = 0 \quad (19)$$

Employing the Pekeris type (approximation 1) (eq.3) and $\alpha = \frac{1}{b}$, Eq. (19) becomes;

$$\frac{d^2 R_{n\ell}(r)}{dr^2} + \frac{1}{(1-qe^{-2\alpha r})^2}\left[\frac{2\mu E_{nl}}{\hbar^2}(1-qe^{-2\alpha r})^2 - \frac{2\mu}{\hbar^2}\left(\left(V_0+\frac{\alpha^2}{2}\right)e^{-\alpha r}(1-qe^{-\alpha r})\right) - \frac{2\mu}{\hbar^2}(V_1 e^{2\alpha r}) - \frac{(D+2\ell-1)(D+2\ell-3)}{4}(\alpha^2 e^{-\alpha r})\right] R_{n\ell}(r) \qquad (20)$$

Eq. (20) can be simplified further by introducing the following dimensionless abbreviations

$$\begin{cases} -\varepsilon_n = \frac{2\mu E_{nl}}{\hbar^2 \alpha^2} \\ \beta = \frac{2\mu V_1}{\hbar^2 \alpha^2} \\ \chi = -\frac{2\mu\left(V_0+\frac{\alpha^2}{2}\right)}{\hbar^2 \alpha^2} \\ \eta = \frac{(D+2\ell-1)(D+2\ell-3)}{4} \end{cases} \qquad (21)$$

And using the transformation $s = e^{-\alpha r}$ so as to enable us apply the NU method as a solution of the hypergeometric type

$$\frac{d^2 R_{n\ell}(r)}{dr^2} = \alpha^2 s^2 \frac{d^2 R_{n\ell}(s)}{ds^2} + \alpha^2 s \frac{d R_{n\ell}(s)}{ds} \qquad (22)$$

$$\frac{d^2 R_{n\ell}(s)}{ds^2} + \frac{1-qs}{s(1-qs)} \frac{dR_{n\ell}(s)}{ds} + \frac{1}{s^2(1-qs)^2}[-s^2(\varepsilon_n q^2 + \chi q - \beta) + s(2\varepsilon_n q + \chi - \eta) - \varepsilon_n]R_{n\ell}(s) = 0$$

(23)

Comparing Eq. (23) and Eq. (6), we have the following parameters

$$\begin{cases} \tilde{\tau}(s) = 1 - qs \\ \sigma(s) = s(1-qs) \\ \tilde{\sigma}(s) = -s^2(\varepsilon_n q^2 + \chi q - \beta) + s(2\varepsilon_n q + \chi - \eta) - \varepsilon_n \end{cases} \quad (24)$$

Substituting these polynomials into Eq. (14), we get $\pi(s)$ to be

$$\pi(s) = -\frac{qs}{2} \pm \sqrt{(a-k)s^2 + (b+k)s + c} \quad (25)$$

where

$$\begin{cases} a = \frac{q^2}{4} + \varepsilon_n q^2 + \chi q - \beta \\ b = -(2\varepsilon_n q + \chi - \eta) \\ c = \varepsilon_n \end{cases} \quad (26)$$

To find the constant $k$, the discriminant of the expression under the square root of Eq. (25) must be equal to zero. As such, we have that

$$k_{\pm} = -(\eta - \chi) \pm 2\sqrt{\varepsilon_n \left(\frac{q^2}{4} + \beta + \eta q\right)} \quad (27)$$

Substituting Eq. (27) into Eq. (25) yields

$$\pi(s) = -\frac{qs}{2} \pm \left[\left(q\sqrt{\varepsilon_n} + \sqrt{\left(\frac{q^2}{4} + \beta + \eta q\right)}\right)s - \sqrt{\varepsilon_n}\right] \quad (28)$$

From the knowledge of NU method, we choose the expression $\pi(s)_-$ which the function $\tau(s)$ has a negative derivative. This is given by

$$k_- = -(\eta - \chi) - 2\sqrt{\varepsilon_n \left(\frac{q^2}{4} + \beta + \eta q\right)} \quad (29)$$

with $\tau(s)$ being obtained as

$$\tau(s) = 1 - 2qs - 2\left[\left(\sqrt{\left(\frac{q^2}{4} + \beta + \eta q\right)} + q\sqrt{\varepsilon_n}\right)s - \sqrt{\varepsilon_n}\right] \quad (30)$$

Referring to Eq. (15), we define the constant $\lambda$ as

$$\lambda = -\frac{q}{2} - \left(q\sqrt{\varepsilon_n} + \sqrt{\left(\frac{q^2}{4} + \beta + \eta q\right)}\right) - (\eta - \chi) - 2\sqrt{\varepsilon_n\left(\frac{q^2}{4} + \beta + \eta q\right)} \qquad (31)$$

Taking the derivative of $\tau(s)$ with respect to $s$ in Eq.(30) we get;

$$\tau'(s) = -2\left(q + \left(\sqrt{\left(\frac{q^2}{4} + \beta + \eta q\right)} + q\sqrt{\varepsilon_n}\right)\right) < 0 \qquad (32)$$

From Eq. (24), taking the derivative of $\sigma(s)$ with respect to $s$, we get;

$$\sigma''(s) = -2q \qquad (33)$$

Substituting Eq. (28) and (29) into Eq. (16) and carrying out simple algebra, we get; $\lambda_n$. Setting; $\lambda_n = \lambda$ and carrying out some algebraic manipulations, we have;

$$\varepsilon_n = \frac{1}{4}\left[\frac{\left(n+\frac{1}{2}+\sqrt{\frac{1}{4}+\frac{\beta}{q^2}+\frac{\eta}{q}}\right)^2 - \frac{\beta}{q^2} - \frac{\chi}{q}}{\left(n+\frac{1}{2}+\sqrt{\frac{1}{4}+\frac{\beta}{q^2}+\frac{\eta}{q}}\right)}\right]^2 \qquad (34)$$

Substituting Eqs. (21) into Eq. (34) yields the energy eigenvalue equation of the generalized shifted Hulthén potential in $D$ dimension in the form

$$E_{n\ell}^{Approx.1} = -\frac{\hbar^2\alpha^2}{8\mu}\left[\frac{\left(n+\frac{1}{2}+\sqrt{\frac{1}{4}+\frac{2\mu V_1}{\hbar^2\alpha^2 q^2}+\frac{(D+2\ell-1)(D+2\ell-3)}{4q}}\right)^2 - \frac{2\mu V_1}{\hbar^2\alpha^2 q^2} + \frac{2\mu\left(V_0+\frac{\alpha^2}{2}\right)}{\hbar^2\alpha^2 q}}{\left(n+\frac{1}{2}+\sqrt{\frac{1}{4}+\frac{2\mu V_1}{\hbar^2\alpha^2 q^2}+\frac{(D+2\ell-1)(D+2\ell-3)}{4q}}\right)}\right]^2 \qquad (35)$$

Again by using approximation (2) and repeat the above procedure, we can consequently obtain the energy eigenvalues as;

$$E_{n\ell}^{Approx.2} = -\frac{\hbar^2\alpha^2}{8\mu}\left[\frac{\left(n+\frac{1}{2}+\sqrt{\frac{1}{4}+\frac{2\mu V_1}{\hbar^2\alpha^2 q^2}+\frac{(D+2\ell-1)(D+2\ell-3)}{4q^2}}\right)^2 - \frac{2\mu V_1}{\hbar^2\alpha^2 q^2} + \frac{2\mu\left(V_0+\frac{\alpha^2}{2}\right)}{\hbar^2\alpha^2 q} + \frac{(D+2\ell-1)(D+2\ell-3)}{4}\left(\frac{e^\alpha}{q}-\frac{1}{q^2}\right)}{\left(n+\frac{1}{2}+\sqrt{\frac{1}{4}+\frac{2\mu V_1}{\hbar^2\alpha^2 q^2}+\frac{(D+2\ell-1)(D+2\ell-3)}{4q^2}}\right)}\right]^2 \qquad (36)$$

Again by using approximation (3) and repeat the above procedure, we can consequently obtain the energy eigenvalues as

$$E_{n\ell}^{Approx.3} = \frac{\hbar^2\alpha^2}{2\mu}\left(\left(\frac{(D+2\ell-1)(D+2\ell-3)C_0}{4}\right) - \frac{1}{4}\left[\frac{\left(n+\frac{1}{2}+\sqrt{\frac{1}{4}+\frac{2\mu V_1}{\hbar^2\alpha^2 q^2}+\frac{(D+2\ell-1)(D+2\ell-3)}{4q}}\right)^2 - \frac{2\mu V_1}{\hbar^2\alpha^2 q^2}+\frac{2\mu\left(V_0+\frac{\alpha^2}{2}\right)}{\hbar^2\alpha^2 q}}{\left(n+\frac{1}{2}+\sqrt{\frac{1}{4}+\frac{2\mu V_1}{\hbar^2\alpha^2 q^2}+\frac{(D+2\ell-1)(D+2\ell-3)}{4q}}\right)}\right]^2\right)$$
(37)

The corresponding wave functions can be evaluated by substituting $\pi(s)_-$ and $\sigma(s)$ from Eq. (28) and Eq. (24) respectively into Eq. (9) and solving the first order differential equation. This gives

$$\Phi(s) = s^{\sqrt{\varepsilon_n}}(1-qs)^{\frac{1}{2}+\sqrt{\frac{1}{4}+\frac{\beta}{q^2}+\frac{\eta}{q}}}$$
(38)

The weight function $\rho(s)$ from Eq. (12) can be obtained as

$$\rho(s) = s^{2\sqrt{\varepsilon_n}}(1-qs)^{2\sqrt{\frac{1}{4}+\frac{\beta}{q^2}+\frac{\eta}{q}}}$$
(39)

From the Rodrigues relation of Eq. (11), we obtain

$$y_n(s) \equiv N_{n,l}P_n^{\left(2\sqrt{\varepsilon_n},\ 2\sqrt{\frac{1}{4}+\frac{\beta}{q^2}+\frac{\eta}{q}}\right)}(1-2qs)$$
(40)

where $P_n^{(\theta,\vartheta)}$ is the Jacobi Polynomial.

Substituting $\Phi(s)$ and $y_n(s)$ from Eq. (32) and Eq. (34) respectively into Eq. (3), we obtain

$$R_{nl}(s) = N_{n,l}\,s^{\sqrt{\varepsilon_n}}(1-qs)^{\frac{1}{2}+\sqrt{\frac{1}{4}+\frac{\beta}{q^2}+\frac{\eta}{q}}}P_n^{\left(2\sqrt{\varepsilon_n},\ 2\sqrt{\frac{1}{4}+\frac{\beta}{q^2}+\frac{\eta}{q}}\right)}(1-2qs)$$
(41)

## 4 Results and Discussion

To show the accuracy of our results, we obtained the eigenvalues (in units of $fm^{-1}$.) numerically (Tables 1-3) for arbitrary quantum numbers $n$ and $l$ with the potential parameter $\alpha = 0.025, 0.050$ and $0.075\ fm^{-1}$ in 3D. In Table 1, we present the numerical results for generalized shifted Hulthén Potential in natural units for $q = 1$ (absence of deformation(generalized shifted Hulthén)), potential strength, $V_0 = 5fm^{-1}$ and $V_1 = 2fm^{-1}$ and $\alpha = 1/b = 0.025, 0.050$ and $0.075fm^{-1}$. The domain in which the screening parameter $\alpha \to 0\ fm^{-1}$ is called the low screening regime. In this regime the generalized shifted Hulthén potential model becomes the constant. For a fixed value of angular momentum quantum $l$, the energy spectrum increases as the principal quantum number $n$ increases for the strong potential coupling strengths, $V_0 = 5$ and $V_1 = 2fm^{-1}$ as seen on Table 1. For a fixed value of angular momentum quantum $l$, the energy spectrum increases as the principal quantum number $n$ increases for a small screening parameter (i.e., low screening regime) "$\alpha$". An increase in angular momentum quantum $l$, leads to an increase in the energy spectrum as the principal quantum number $n$ increases for a varying screening parameter $\alpha$ and for a strong potential coupling strength, ($V_0$ and $V_1$). For a weak potential coupling strength, ($V_0$ and $V_1$), solutions are ignored due to the presence of imaginary terms and the energy spectrum is not complex but real. Beyond this, we can observe from Table 1, the energy eigenvalue is strongly bounded and an increase in rotational quantum number $l$ makes energy become more attractive (i.e., the energy becomes more negative) with increasing $\alpha$.

Interestingly, the above observation is the same in the presence of the deformation parameter in the system as shown in Table 2 and 3 for $q = 2$ and $q = -2$ except for the fact that the presence of the deformation parameter makes the energy become more attractive. Although the energy is more attractive when the deformation parameter is less than 0 ($q < 0$). The analytical expressions for the total energy levels of this system is found to be general in the sense that it is obtained in arbitrary dimensions and the presence of the deformation parameter provides an avenue to arrive at special cases e.g. when $q \to -q$, we arrive at the Wood-Saxon potential etc.

In this study, three approximation schemes were employed. To show that Eqs.(3-5) are good approximation scheme we compared $\frac{1}{r^2}$ and the approximation scheme with $\alpha = 0.025$ in Fig. 1 for $q = 1$. In Fig.2 the variation of shifted Hulthén potential, special Hulthén potential and Hulthén potential with $r$ for $V_0 = 5, V_1 = 2$ and $\alpha = 0.025$ was plotted. This was done in order to enable us show the behaviour of the shifted Hulthén potential. It can be easily observed that the Hulthén and special Hulthén behave in the same manner. Figs. (3-6) shows the behavior of the wave function in the presence and absence of the deformation parameter. Finally, we point out that these exact results obtained for this newly proposed form of the potential (2) may have some interesting applications in the study of different quantum mechanical systems, atomic and molecular physics.

## 5. SPECIAL CASE

In this section, we make some adjustments of constants in Eq. (2) and Eqs. (35), (36) and (37) to have the following cases:

First, we study the s-wave case ($\ell = 0$) for $D = 3$ and $q = 1$. The solutions of energy eigenvalues Eqs. (35), (36) and (37), reduce to the following equation

$$E_{n\ell}^{Approx.1} = E_{n\ell}^{Approx.2} = E_{n\ell}^{Approx.3} = -\frac{\hbar^2\alpha^2}{8\mu}\left[\frac{\left(n+\frac{1}{2}+\sqrt{\frac{1}{4}+\frac{2\mu V_1}{\hbar^2\alpha^2}}\right)^2 - \frac{2\mu V_1}{\hbar^2\alpha^2} + \frac{2\mu\left(V_0+\frac{\alpha^2}{2}\right)}{\hbar^2\alpha^2}}{\left(n+\frac{1}{2}+\sqrt{\frac{1}{4}+\frac{2\mu V_1}{\hbar^2\alpha^2}}\right)}\right]^2 \qquad (42)$$

### 5.1 HULTHÉN POTENTIAL

If we set $V_1 = 0, V^1 = \frac{1}{2b^2} = 0$, $= \frac{1}{b}$, $q = 1$ and $V_0 = -V_0$ in Eq.(2). We obtain the Hulthén potential as follows;

$$V(r) = \frac{-V_0 e^{-\alpha r}}{1-e^{-\alpha r}} \qquad (43)$$

Its energy eigenvalue equation can be deduced from Eqs. (34), (36) and (37) as

$$E_{n\ell}^{Approx.1} = -\frac{\hbar^2\alpha^2}{8\mu}\left[\frac{\left(n+\frac{1}{2}+\sqrt{\frac{1}{4}+\frac{(D+2\ell-1)(D+2\ell-3)}{4}}\right)^2 - \frac{2\mu V_0}{\hbar^2\alpha^2}}{\left(n+\frac{1}{2}+\sqrt{\frac{1}{4}+\frac{(D+2\ell-1)(D+2\ell-3)}{4}}\right)}\right]^2 \qquad (44)$$

$$E_{n\ell}^{Approx.2} = -\frac{\hbar^2\alpha^2}{8\mu}\left[\frac{\left(n+\frac{1}{2}+\sqrt{\frac{1}{4}+\frac{(D+2\ell-1)(D+2\ell-3)}{4}}\right)^2 - \frac{2\mu V_0}{\hbar^2\alpha^2} + \frac{(D+2\ell-1)(D+2\ell-3)}{4}(e^\alpha-1)}{\left(n+\frac{1}{2}+\sqrt{\frac{1}{4}+\frac{(D+2\ell-1)(D+2\ell-3)}{4}}\right)}\right]^2 \qquad (45)$$

and

$$E_{n\ell}^{Approx.3} = \frac{\hbar^2\alpha^2}{2\mu}\left(\left(\frac{(D+2\ell-1)(D+2\ell-3)C_0}{4}\right) - \frac{1}{4}\left[\frac{\left(n+\frac{1}{2}+\sqrt{\frac{1}{4}+\frac{(D+2\ell-1)(D+2\ell-3)}{4}}\right)^2 - \frac{2\mu V_0}{\hbar^2\alpha^2}}{\left(n+\frac{1}{2}+\sqrt{\frac{1}{4}+\frac{(D+2\ell-1)(D+2\ell-3)}{4}}\right)}\right]^2\right) \qquad (46)$$

In 3D, Eqs. (44),(45) and (46) reduce to;

$$E_{n\ell}^{Approx.1} = -\frac{\hbar^2\alpha^2}{8\mu}\left[\frac{\left(n+\frac{1}{2}+\sqrt{\frac{1}{4}+\ell(\ell+1)}\right)^2 - \frac{2\mu V_0}{\hbar^2\alpha^2}}{\left(n+\frac{1}{2}+\sqrt{\frac{1}{4}+\ell(\ell+1)}\right)}\right]^2 \qquad (47)$$

$$E_{n\ell}^{Approx.2} = -\frac{\hbar^2\alpha^2}{8\mu}\left[\frac{\left(n+\frac{1}{2}+\sqrt{\frac{1}{4}+\ell(\ell+1)}\right)^2 - \frac{2\mu V_0}{\hbar^2\alpha^2} + \ell(\ell+1)(e^\alpha-1)}{\left(n+\frac{1}{2}+\sqrt{\frac{1}{4}+\ell(\ell+1)}\right)}\right]^2 \qquad (48)$$

and

$$E_{n\ell}^{Approx.3} = \frac{\hbar^2\alpha^2}{2\mu}\left((\ell(\ell+1)C_0) - \frac{1}{4}\left[\frac{\left(n+\frac{1}{2}+\sqrt{\frac{1}{4}+\ell(\ell+1)}\right)^2 - \frac{2\mu V_0}{\hbar^2\alpha^2}}{\left(n+\frac{1}{2}+\sqrt{\frac{1}{4}+\ell(\ell+1)}\right)}\right]^2\right) \qquad (49)$$

Eq. (44) is identical with the energy eigenvalue equation given in Eq. (30) of ref. [51]. More so, if we set $D = 3$, we arrive at the energy eigenvalue equation for the Hulthen potential in $3D$

Eq. (47) is identical with the energy eigenvalues formula given in Eq. (31) of ref. [51]. Eq. (32) of ref. [52], Eq. (24) of ref. [53] and Eq. (28) of ref. [54] and Eq. (36) of ref. [55]. Eq. (49) is identical with the energy eigenvalues formula (34) of [55]

Furthermore, for s-wave ( $\ell = 0$) states, equations (47), (48) and (49) reduce to

$$E_{n\ell}^{Approx.1} = E_{n\ell}^{Approx.2} = E_{n\ell}^{Approx.3} = -\frac{\hbar^2\alpha^2}{8\mu}\left[\frac{(n+1)^2 - \frac{2\mu V_0}{\hbar^2\alpha^2}}{(n+1)}\right]^2 \qquad (50)$$

which is identical to the ones obtained before using the factorization method [56], SUSYQM approach [57-59], NU method [60,54,55] and AIM Eq.(39) of Ref.[36].

### 5.2 WOODS-SAXON POTENTIAL

If $V_1 = 0, V^1 = \frac{1}{2b^2} = 0, = \frac{1}{b}$, and $q = -1$ in Eq. (2), we can obtain the Woods-Saxon potential of the form;

$$V(r) = \frac{-V_0 e^{-\alpha r}}{1+e^{-\alpha r}} \qquad (51)$$

Its energy eigenvalue equation can be deduced from Eqs. (34), (36) and (37) as

$$E_{n\ell}^{Approx.1} = -\frac{\hbar^2\alpha^2}{8\mu}\left[\frac{\left(n+\frac{1}{2}+\sqrt{\frac{1}{4}-\frac{(D+2\ell-1)(D+2\ell-3)}{4}}\right)^2 - \frac{2\mu V_0}{\hbar^2\alpha^2}}{\left(n+\frac{1}{2}+\sqrt{\frac{1}{4}-\frac{(D+2\ell-1)(D+2\ell-3)}{4}}\right)}\right]^2 \qquad (52)$$

Again by using approximation (2) and repeat the above procedure, we can consequently obtain the energy eigenvalues as;

$$E_{n\ell}^{Approx.2} = -\frac{\hbar^2\alpha^2}{8\mu}\left[\frac{\left(n+\frac{1}{2}+\sqrt{\frac{1}{4}+\frac{(D+2\ell-1)(D+2\ell-3)}{4}}\right)^2 - \frac{2\mu V_0}{\hbar^2\alpha^2} - \frac{(D+2\ell-1)(D+2\ell-3)}{4}(e^\alpha+1)}{\left(n+\frac{1}{2}+\sqrt{\frac{1}{4}+\frac{(D+2\ell-1)(D+2\ell-3)}{4}}\right)}\right]^2 \qquad (53)$$

Again by using approximation (3) and repeat the above procedure, we can consequently obtain

the energy eigenvalues as

$$E_{n\ell}^{Approx.3} = \frac{\hbar^2\alpha^2}{2\mu}\left(\left(\frac{(D+2\ell-1)(D+2\ell-3)C_0}{4}\right) - \frac{1}{4}\left[\frac{\left(n+\frac{1}{2}+\sqrt{\frac{1}{4}-\frac{(D+2\ell-1)(D+2\ell-3)}{4}}\right)^2 - \frac{2\mu V_0}{\hbar^2\alpha^2}}{\left(n+\frac{1}{2}+\sqrt{\frac{1}{4}-\frac{(D+2\ell-1)(D+2\ell-3)}{4}}\right)}\right]^2\right) \quad (54)$$

Eqs.(52),(53) and (54) are the energy equation for Woods-Saxon potential in $D$ Dimensions with different approximation scheme. If $D = 3$, eqs.(52),(53) and (54) reduces to energy equation for Woods-Saxon potential in 3D. More so, Eqs. (52) is in agreement with Eq. (30) of ref.[61] and Eq.(64) of [62,63]

## 6.0 CONCLUSION

In this work, we have studied the bound state solutions of the Schrodinger equation with generalized shifted Hulthén Potential in $D$ dimensions using NU method. We used three different approximation scheme to deal with the centrifugal term, we obtain the energy eigenvalues and the corresponding eigenfunctions and also discussed some special cases of the potential. We have calculated numerical energy eigenvalues and presented plots for various values of the potential parameters. It is found out that the numerical values were in good agreement. The results are in excellent agreement with literature. Finally, our results can find many applications in quantum mechanical systems, atomic and molecular physics.

## ACKNOWLEDGMENTS

C. O. Edet dedicates this work to his Late Father. In addition, C. O. Edet acknowledges Dr. A. N. Ikot for his continuous supports and encouragement.

Table 1. The bound state energy levels (in units of fm−1) of the generalized shifted Hulthén Potential for various values of $n$, $l$ and for $\hbar = \mu = 1, q = 1$, $V_0 = 5, V_1 = 2$ and $1/b = 0.025, 0.050$ and $0.075$

| $n$ | $l$ | $1/b$ | Approx. 1 | Approx. 2 | Approx. 3 |
|---|---|---|---|---|---|
| 0 | 1 | 0.025 | -3.117514261 | -3.11753389 | -3.117462178 |
|   |   | 0.050 | -3.110700106 | -3.110857879 | -3.110491773 |
|   |   | 0.075 | -3.104584691 | -3.10511967 | -3.104115941 |
| 1 | 1 | 0.025 | -3.103082968 | -3.10310231 | -3.103030885 |
|   |   | 0.050 | -3.084157322 | -3.084310638 | -3.083948989 |
|   |   | 0.075 | -3.06815173 | -3.068664704 | -3.06768298 |
| 2 | 1 | 0.025 | -3.089786861 | -3.089805928 | -3.089734778 |
|   |   | 0.050 | -3.061877318 | -3.062026488 | -3.061668985 |
|   |   | 0.075 | -3.040735741 | -3.041228935 | -3.040266991 |
| 0 | 2 | 0.025 | -3.117139376 | -3.117198239 | -3.116983126 |
|   |   | 0.050 | -3.109268049 | -3.109740695 | -3.108643049 |
|   |   | 0.075 | -3.101523451 | -3.103123335 | -3.100117201 |
| 1 | 2 | 0.025 | -3.102737046 | -3.102795052 | -3.102580796 |
|   |   | 0.050 | -3.082946589 | -3.083405908 | -3.082321589 |
|   |   | 0.075 | -3.065803052 | -3.067337454 | -3.064396802 |
| 2 | 2 | 0.025 | -3.089468696 | -3.089525875 | -3.089312446 |
|   |   | 0.050 | -3.060870455 | -3.061317376 | -3.060245455 |
|   |   | 0.075 | -3.039019378 | -3.040494886 | -3.037613128 |
| 0 | 3 | 0.025 | -3.116578656 | -3.116696314 | -3.116266156 |
|   |   | 0.050 | -3.107144798 | -3.108088059 | -3.105894798 |
|   |   | 0.075 | -3.09705253 | -3.100237354 | -3.09424003 |
| 1 | 3 | 0.025 | -3.102219691 | -3.102335639 | -3.101907191 |
|   |   | 0.050 | -3.08115308 | -3.082069834 | -3.07990308 |
|   |   | 0.075 | -3.062385156 | -3.065440549 | -3.059572656 |
| 2 | 3 | 0.025 | -3.088992914 | -3.089107212 | -3.088680414 |
|   |   | 0.050 | -3.059380745 | -3.060272836 | -3.058130745 |
|   |   | 0.075 | -3.036536508 | -3.039475458 | -3.033724008 |
| 0 | 4 | 0.025 | -3.115834011 | -3.116029961 | -3.115313178 |
|   |   | 0.050 | -3.104359635 | -3.105927274 | -3.102276302 |
|   |   | 0.075 | -3.091311559 | -3.096587012 | -3.086624059 |
| 1 | 4 | 0.025 | -3.10153273 | -3.101725838 | -3.101011897 |
|   |   | 0.050 | -3.078803444 | -3.080327222 | -3.076720111 |
|   |   | 0.075 | -3.058019492 | -3.063082576 | -3.053331992 |
| 2 | 4 | 0.025 | -3.088361249 | -3.088551611 | -3.087840416 |
|   |   | 0.050 | -3.057432485 | -3.058915442 | -3.055349152 |
|   |   | 0.075 | -3.033393131 | -3.038265041 | -3.028705631 |
| 3 | 4 | 0.025 | -3.076273284 | -3.076461004 | -3.075752451 |
|   |   | 0.050 | -3.039925015 | -3.04136992 | -3.037841682 |
|   |   | 0.075 | -3.016497519 | -3.021196724 | -3.011810019 |

Table 2. The bound state energy levels (in units of fm−1) of the generalized shifted Hulthén Potential for various values of $n$, $l$ and for $\hbar = \mu = 1$, $q = 2$, $V_0 = 5$, $V_1 = 2$ and $\alpha = 0.025$

| $n$ | $l$ | Approx. 1 | Approx. 2 | Approx. 3 |
| --- | --- | --- | --- | --- |
| 0 | 1 | -3.07804408 | -3.07900825 | -3.077991997 |
| 1 | 1 | -2.991656822 | -2.992562348 | -2.991604739 |
| 2 | 1 | -2.911896391 | -2.912748234 | -2.911844308 |
| 3 | 1 | -2.838174895 | -2.83897748 | -2.838122812 |
| 0 | 2 | -3.075800058 | -3.07868871 | -3.075643808 |
| 1 | 2 | -2.989586095 | -2.992299144 | -2.989429845 |
| 2 | 2 | -2.909983401 | -2.912535695 | -2.909827151 |
| 3 | 2 | -2.836405828 | -2.838810616 | -2.836249578 |
| 0 | 3 | -3.072445134 | -3.078210908 | -3.072132634 |
| 1 | 3 | -2.986490142 | -2.991905695 | -2.986177642 |
| 2 | 3 | -2.907123184 | -2.912218111 | -2.906810684 |
| 3 | 3 | -2.833760719 | -2.838561429 | -2.833448219 |
| 0 | 4 | -3.067992528 | -3.077576635 | -3.067471695 |
| 1 | 4 | -2.982381021 | -2.991383619 | -2.981860188 |
| 2 | 4 | -2.903326769 | -2.911796942 | -2.902805936 |
| 3 | 4 | -2.830249678 | -2.838231239 | -2.829728845 |

Table 3. The bound state energy levels (in units of fm−1) of the generalized shifted Hulthén Potential for various values of $n$, $l$ and for $\hbar = \mu = 1$, $q = -2$, $V_0 = 5$, $V_1 = 2$ and $\alpha = 0.025$

| $n$ | $l$ | Approx. 1 | Approx. 2 | Approx. 3 |
| --- | --- | --- | --- | --- |
| 0 | 1 | -3.020072465 | -3.017282372 | -3.020020382 |
| 1 | 1 | -2.817314959 | -2.814748239 | -2.817262876 |
| 2 | 1 | -2.629306878 | -2.626941798 | -2.629254795 |
| 3 | 1 | -2.454721581 | -2.452538949 | -2.454669498 |
| 0 | 2 | -3.025345822 | -3.016966052 | -3.025189572 |
| 1 | 2 | -2.822200951 | -2.814492116 | -2.822044701 |
| 2 | 2 | -2.633840768 | -2.626737579 | -2.633684518 |
| 3 | 2 | -2.458934669 | -2.452379472 | -2.458778419 |
| 0 | 3 | -3.033281168 | -3.016493064 | -3.032968668 |
| 1 | 3 | -2.829553036 | -2.814109248 | -2.829240536 |
| 2 | 3 | -2.640662724 | -2.626432419 | -2.640350224 |
| 3 | 3 | -2.46527366 | -2.452141296 | -2.46496116 |
| 0 | 4 | -3.043909145 | -3.01586518 | -3.043388312 |
| 1 | 4 | -2.839399168 | -2.813601206 | -2.838878335 |
| 2 | 4 | -2.649798308 | -2.626027709 | -2.649277475 |
| 3 | 4 | -2.473761971 | -2.45182566 | -2.473241138 |

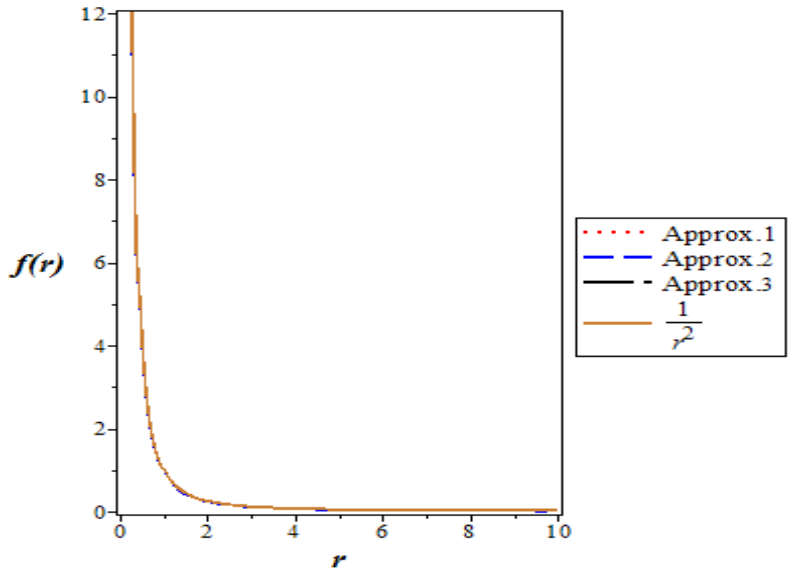

**Figure 1**; Comparison between $1/r^2$ and the approximation scheme as functions of $r$ for $\alpha = 0.025$

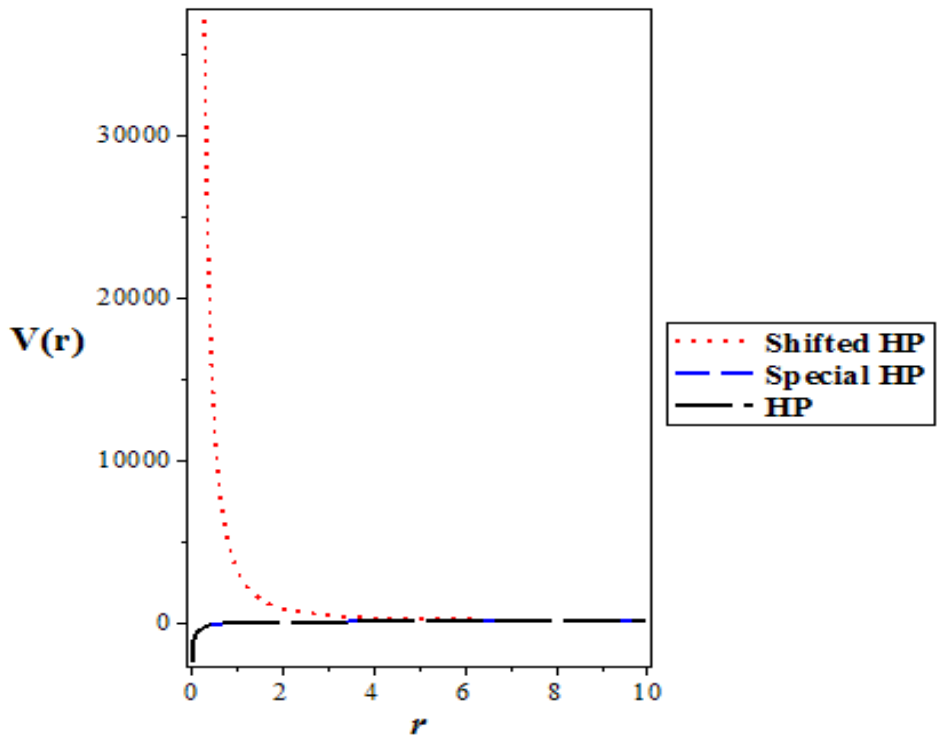

**Figure 2**; Variation of shifted Hulthén Potential, Special Hulthén Potential and Hulthén Potential with $r$ for $V_0 = 5, V_1 = 2$ and $\alpha = 0.025$

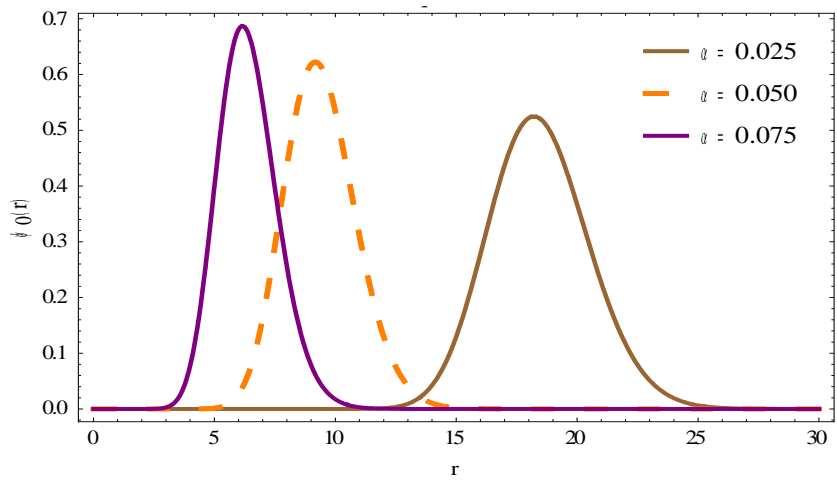

**Figure 3**: Wave functions against radial distance for different values of the screening parameter $(\alpha)$. We chose $n = \ell = 0, q = 1, V_0 = 5$ and $V_1 = 2$

.

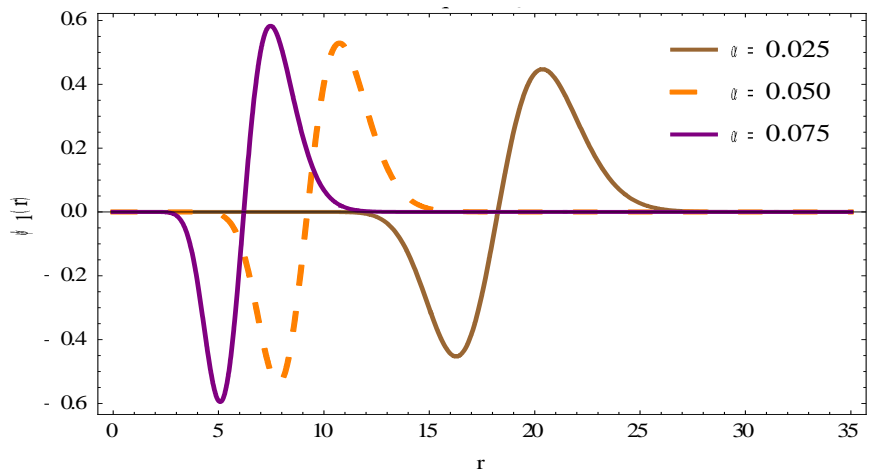

**Figure 4**: Wave functions against radial distance for different values of the screening parameter $(\alpha)$. $n = 1, \ell = 0, q = 1, V_0 = 5$ and $V_1 = 2$

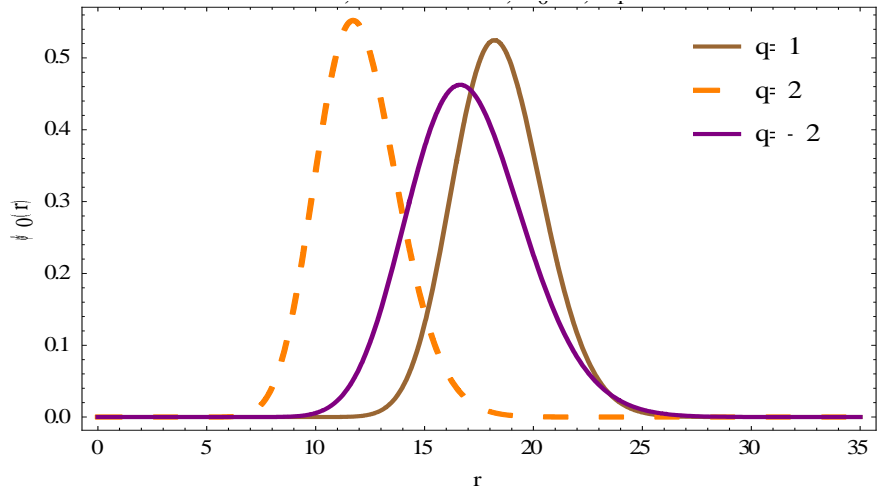

**Figure 5**: Wave functions against radial distance for different values of the deformation parameter $(q)$. We chose $n = \ell = 0$, $\alpha = 0.025$, $V_0 = 5$ and $V_1 = 2$.

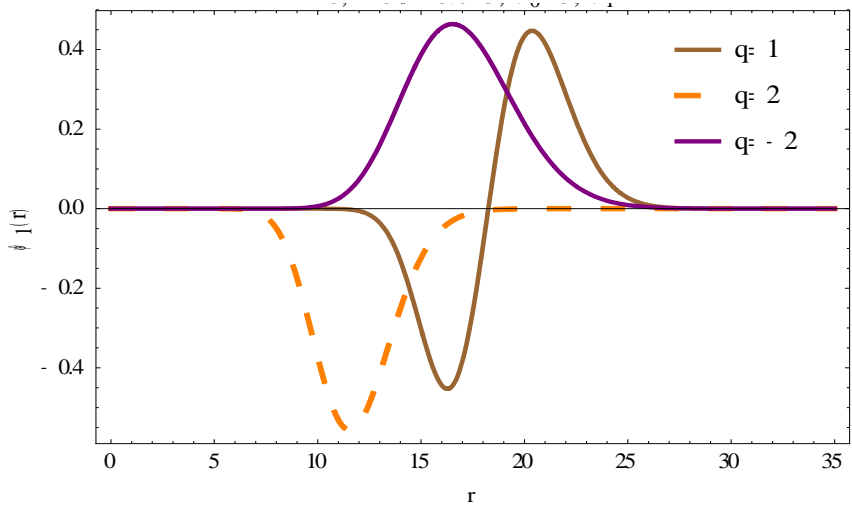

**Figure 6**: Wave functions against radial distance for different values of the deformation parameter $(q)$. We chose $n = 1, \ell = 0$, $\alpha = 0.025$, $V_0 = 5$ and $V_1 = 2$.